\documentclass[bimj,fleqn]{w-art}
\usepackage{times}
\usepackage{w-thm}
\usepackage[authoryear]{natbib}
\usepackage{amssymb, amsmath}

\newcommand{\bbeta}{\boldsymbol{\beta}}
\newcommand{\bgamma}{\boldsymbol{\gamma}}
\newcommand{\bgammastar}{\boldsymbol{\gamma}^{\star}}

\newcommand{\argmax}{\mbox{arg\,max}} %% attention cetait un DeclareOperator

\newcommand{\BIC}{\mathbf{BIC}}

\newcommand{\bx}{\boldsymbol{x}}
\newcommand{\by}{\mathbf{y}}
\newcommand{\bX}{\textbf{x}}
\newcommand{\bOG}{\boldsymbol{\Omega}_{\boldsymbol{\gamma}}}
\newcommand{\Prob}{\mathbb{P}}

\def\bSig\mathbf{\Sigma}

\usepackage[]{graphicx}
\usepackage{algorithmic}
\usepackage{algorithm}

\floatname{algorithm}{Algorithm}
\usepackage[toc,page]{appendix}

\begin{document}
%%    The information for the title page will be placed between
%%    \begin{document} and \maketitle. The order of most entries
%%    is determined by the class file and can not be changed by
%%    rearranging them. The maketitle command follows after the
%%    abstract.
%%
%%    Most of the following commands will be completed by the publisher.
%%
%%    The copyrightyear is defined in the .clo file as the first argument
%%    of the copyrightinfo command. If the copyrightyear differs from that
%%    value it might be adjusted by the following definition:
%%
%% \renewcommand{\copyrightyear}{2004}% uncomment to change the copyrightyear.
%%
\DOIsuffix{bimj.DOIsuffix}
%%
%% issueinfo for the header line
\Volume{46}
\Issue{1}
\Year{2004}
%%
%%    First and last pagenumber of the article. If the option
%%    'autolastpage' is set (default) the second argument may be left empty.
\pagespan{1}{}
%%
%%    Dates will be filled in by the publisher. The 'reviseddate' and
%%    'dateposted' (Published online) entry may be left empty.
\Receiveddate{XX XX XX}
\Reviseddate{XX XX XX}
\Accepteddate{XX XX XX}
\Dateposted{XX XX XX}

%% \pretitle{Editor's Choice}

%% We have a short and a long form for the title. The short form
%% (optional argument) goes into the running head.

\title[]{Bayesian model selection in logistic regression for the detection of adverse drug reactions}

%% Please do not enter footnotes or \inst{}-notes into the optional
%% argument of the author command. The optional argument will go into
%% the header.  If there is only one address the marker \inst{x} may be
%% omitted.

%% Information for the first author.
\author[M. Marbac]{Matthieu Marbac\inst{1}} \address[\inst{1}]{UMR 1181 B2PHI Inserm, Institut-Pasteur and Universit\'e 
Versailles St Quentin.}
%%
%%    Information for the second author
\author[P. Tubert-Bitter]{Pascale Tubert-Bitter\inst{1}}

%%
%%    Information for the third author
\author[M. Sedki]{Mohammed Sedki\footnote{Corresponding
     author: e-mail: {\sf mohammed.sedki@u-psud.fr}, Phone: +33\,145\,595\,214}\inst{1,2}}
\address[\inst{2}]{Universit\'e. Paris-Sud.}
%%
%%    \dedicatory{This is a dedicatory.}
\begin{abstract}
\textit{Motivation:}
Spontaneous adverse event reports have an high potential for detecting adverse
drug reactions. However, due to their dimension, the analysis of such databases requires
statistical methods. In this context, disproportionality measures can be used.
Their main idea is to project the data onto contingency tables in order to measure the strength of associations between drugs and adverse events. However, due to the data projection, these methods are sensitive to the problem of co-prescriptions and masking effects. Recently,
logistic regressions have been used with a Lasso type penalty to perform the
detection of associations between drugs and adverse events. On different examples, this approach limits the drawbacks of the disproportionality methods, but the
choice of the penalty value is open to criticism while it strongly influences
the results.
\textit{Results:}
In this paper, we propose to use a logistic regression whose sparsity is
viewed as a model selection challenge. Since the model space is huge, a
Metropolis-Hastings algorithm carries out the model selection by maximizing
the BIC criterion. Thus, we avoid the calibration of penalty or threshold.
During our application on the French pharmacovigilance database, the proposed
method is compared to well established approaches on a reference data set, and
obtains better rates of positive and negative controls. However, many signals (\emph{i.e.} specific drug-event associations) are not detected by the proposed method. So, we conclude that this method
should be used in parallel to existing measures in pharmacovigilance.

\textit{Availability:} 
Code implementing the proposed method is available in R on request from the corresponding author.
\end{abstract}
%% maketitle must follow the abstract.
\maketitle                   % Produces the title.

%% If there is not enough space inside the running head
%% for all authors including the title you may provide
%% the leftmark in one of the following three forms:

%% \renewcommand{\leftmark}
%% {First Author: A Short Title}

%% \renewcommand{\leftmark}
%% {First Author and Second Author: A Short Title}

%% \renewcommand{\leftmark}
%% {First Author et al.: A Short Title}

%% \tableofcontents  % Produces the table of contents.
%\section{First section}
%\subsection{First subsection}

\section{Introduction}
To obtain approval, drugs go through many premarket safety
tests, but adverse drug reactions may not be detected during
these experiments. Many national or international regulatory agencies
have thus introduced pharmacovigilance systems collecting
  spontaneously reported adverse events. Post-approval drug safety
surveillance relies on these reported cases for suspecting that some drugs induce adverse events. They provide huge binary databases that describe each
individual by its drug consumption and its adverse events. Although
spontaneous reporting systems suffer from many biases \citep{Alm07},
they have permitted early identification of associations between drugs and
adverse events \citep{Sza02}. In order to assist pharmacovigilance
experts in managing such databases, statistical methods aiming to put the light on unexpected associations have been proposed.

The most classical methods are based on \emph{disproportionality measures} and
use data projections onto \emph{contingency tables}. Among them, the most popular are: the Proportional Reporting Ratio \citep{Eva01}, the Reporting
Odds Ratio \citep{Van02}, the Bayesian Confidence Propagation Neural Network
\citep{Bat98} and the Gamma Poisson Shrinkage \citep{Dum99}. All of these
methods use a specific statistic which requires a threshold for
detecting associations between drugs and adverse events. The
disproportionality measure is computed for each drug-event pair in the
database and compared to the threshold. Moreover, the data projections onto
the contingency tables provide good computational performances. However, these
projections involve some weakness against the problems of co-prescriptions and
masking effects from highly reported associations for some drugs
\citep{Cas10}. None of these methods is defined as the reference approach. Due
to the shortage of the gold standard sets, their comparison remains a
challenging issue.

The shrinkage \emph{logistic regression} is an interesting alternative to the
methods based on data projections onto contingency tables. In this spirit,
\citet{Cas10} propose to model the probability of an adverse event
conditionally on the drug consumptions by a sparse logistic regression whose 
sparsity is imposed by a \emph{Lasso type penalty} \citep{Tib96}. In this
context, drug $j$ and adverse event $h$ are claimed to be associated when the
coefficient related to drug $j$ in the regression of adverse event $h$ is
strictly positive --- since, in this case, the adverse event occurs more often with the consumption of this drug. However, the choice of the penalty value is a crucial and
very difficult task. Indeed, the penalty value directly influences the signal
detection. \citet{Cas10} propose to use the same penalty for all the
regressions. Moreover, they set the penalty value in order to obtain the same number of
signals as a disproportionality method. A more rigorous method, but more
computationally demanding, could consist in setting the penalty value by
cross-validation where the penalty is set for minimizing the
misclassification error. However, as shown during our numerical application,
 this approach obtains poor results notably due to the database sparsity. 
Recently, \citet{Har13} have used a full logistic regression in a two-step
procedure where the first step consists in empirically selecting a subset of
candidate drugs.

In this paper, the signal detection is performed by a \emph{model selection}
step which avoids the use of any threshold or the calibration of the penalty. In
this context, a model of a logistic regression determines the coefficients
which are not zero. In a Bayesian framework, the best model has the highest
posterior probability but this amount is not explicit. It is also useful to
approximate its logarithm by the Bayesian Information Criterion \citep{Sch78}.
Therefore, the signal detection consists in selecting the model which
maximizes the BIC criterion. Unfortunately, the number of competing models is too huge for
applying an exhaustive approach which computes the BIC criterion
for each competing model. Therefore, the model selection is carried out by a
\emph{Metropolis-Hastings algorithm} \citep{Rob04} which performs a random
walk through the models of interest. This algorithm is classically used for finding the maximum of a function even on a discrete space. In our context, the mode of its
stationary distribution corresponds to the model maximizing the BIC criterion.
Thus, we were able to develop an efficient algorithm by taking advantage of
some features of the data.

In this paper, we compare our model-based procedure to the four disproportionality
methods implemented in the R package \emph{PhViD} and to the Lasso logistic regression implemented in the R package \emph{glmnet}. We use the database arisen from
the French pharmacovigilance which received roughly 20,000 suspected adverse
drug reactions per year from 2000 to 2010. Comparison between
pharmacovigilance procedures is a difficult task. In this paper, we focus on
the four adverse events described in the Observational Medical Outcomes
Partnership (OMOP) reference set \citep{Rya13} and on their 145 relating
drugs. To our knowledge, it is the only reference set recently formed with positive and negative controls to address the
issue of methods assessment in pharmacovigilance.

This article is organised as follows. Section~\ref{sec::method} presents the
parsimonious version of the logistic regression. Section~\ref{sec::MH}
introduces the Metropolis-Hastings algorithm devoted to the model selection.
Section~\ref{sec::application} compares the proposed method to four
disproportionality methods and to the Lasso logistic regression.
Section~\ref{sec::discussion} discusses the limitations and scope of the
proposed approach.

\section{Parsimonious logistic regression} 
\label{sec::method}

\subsection{Spontaneous reporting database}

Spontaneous reporting databases describe $n$ individuals by their
consumptions of $p$ drugs and by the presence or absence of $d$
adverse events. For the purpose of logistic regression, in this
article, we consider one adverse event at a time that we denote by the
binary vector $\by=(y_1,\ldots,y_n)\in \mathbb{B}^n$ where
$\mathbb{B}=\{0,1\}$. More specifically, $y_i=1$ if individual $i$
suffers from this adverse event and $y_i=0$ otherwise. In the
regression context, explanatory variables $\bX=(\bx_1,\ldots,\bx_n)$ indicate the presence or
the absence of drug consumptions. Binary vector $\bx_i=(x_{i1},\ldots,x_{ip})\in \mathbb{B}^p$ indicates the drug consumption of individual $i$ since $x_{ij}=1$ if individual $i$ takes
drug $j$ and $x_{ij}=0$ otherwise.

 \subsection{Logistic regression}
 The probability of the adverse event given the drug
 consumption is assumed to follow a \emph{logit regression}. Model
 $\bgamma=(\gamma_1,\ldots,\gamma_p)\in\mathbb{B}^p$ defines which drugs
 influence the appearance of the adverse event, since $\gamma_j=1$ if
 the coefficient of the regression related to drug $j$ is unconstrained (\emph{i.e.} defined on $\mathbb{R}$) while
 $\gamma_j=0$ if this coefficient is zero. 
The indices of the drugs having a non-zero (respectively zero) coefficient are grouped into the set   $\mathcal{D}_{\bgamma}=\{j : \gamma_{j}=1\}$ (respectively $\mathcal{D}_{\bgamma}^c=\{j : \gamma_j=0\}$).

For model $\bgamma$, the \emph{logit} relationship is
 \begin{equation}
 \label{eq:logit-reg}
   \ln \frac{\Prob(y_i = 1 \mid \bx_i, \bgamma, \bbeta)}{1-\Prob(y_i = 1 \mid \bx_i, \bgamma, \bbeta)}
 = \beta_0 + \sum_{j \in \mathcal{D}_{\bgamma}}\beta_j x_{ij}, 
  \end{equation}
  $\bbeta =
  \big(\beta_0, \beta_1, \ldots, \beta_p\big) \in \bOG$ being the vector
  of regression coefficients for which many coefficients are constrained 
  by $\bgamma$ to be zero, since
 \begin{equation}
 \bOG=\left\{\bbeta \in \mathbb{R}^{p+1}: \forall j \in \mathcal{D}_{\bgamma}^c,\; \beta_j=0\right\}.
 \end{equation} 
Thus, the drugs suspected to induce the adverse event are those belonging to  $\mathcal{D}_{\bgamma}$ and having a positive coefficient in the regression (\emph{i.e.} $\beta_j>0$).
 
Assuming that spontaneous reports consists of $n$ i.i.d. observations, the \emph{adverse event log-likelihood} related to model $\bgamma$ is written as
  \begin{equation}
    \label{eq:log-lik}
    \ell_n\big(\by \mid \bX,\bgamma, \bbeta  \big) = 
    \sum^n_{i = 1} y_i\big(\beta_0 + \sum_{j \in \mathcal{D}_{\bgamma}}\beta_j
    x_{ij}\big) - \ln \Big[1 + \exp\big(\beta_0 + \sum_{j \in \mathcal{D}_{\bgamma}}\beta_j
    x_{ij}\big)\Big].
   \end{equation}
   Obviously, the indices of $\bx_i$ impacting the log-likelihood
   value are those belonging to $\mathcal{D}_{\bgamma}$.  In practice,
   it is often more numerically efficient to compute the adverse event
   log-likelihood by using the unique profiles of observations
   impacting the likelihood.  This weighted form of the log-likelihood
   is described in Appendix~A.
  
   From the database, the Maximum Likelihood Estimates (MLE)
   $\widehat{\bbeta}_{\bgamma}$ is defined by
 \begin{equation}
 \label{eq:MLE}
 \widehat{\bbeta}_{\bgamma}= \argmax_{\bbeta \in \bOG} \ell_n\big(\by\mid \bX, \bgamma, \bbeta \big).
 \end{equation}
 To assess~\eqref{eq:MLE}, we need to solve the derivative likelihood
 equations using the classical Newton-Raphson method (see~\cite{Noc06}).
 However, the MLE  is well defined only if the overlapping 
 conditions of~\cite{Sil81} are satisfied %. More precisely, 
% fix one drug $j$ such as $j\in\mathcal{D}_{\bgamma}$, the conditions are
 %\begin{equation}
 %  l_{0j} < u_{1j} \quad \text{and} \quad l_{1j} < u_{0j}, \label{eq:prevcond}
 %\end{equation}
 %where $l_{0j} = \min\{x_{ij} \mid y_i = 0\}$, $u_{0j} = \max\{x_{ij} \mid y_i
 %= 0\}$, $l_{1j} = \min\{x_{ij} \mid y_i = 1\}$ and $u_{1j} = \max\{x_{ij}
 %\mid y_i = 1\}$. These conditions are equivalent to have a positive length
 %overlap of the intervals $[l_{0j}, u_{0j}]$ and $[l_{1j}, u_{1j}]$ in the
 %scalar case
   (see also the discussion of ~\cite{Owen14}). % In spontaneous reporting
% database settings, $x_{ij} \in \mathbb{B}$ and then \eqref{eq:prevcond} is equivalent to
Thus, for the binary variables, the MLE  is well defined only if
 \begin{equation}
 \label{eq::cond}
 \forall (h_y,h_x) \in \mathbb{B}^2,\; \forall j \in \mathcal{D}_{\bgamma},\; \exists i \in \mathcal{I}_{h_y}: x_{ij}=h_x,
 \end{equation}
 where $\mathcal{I}_{h_y}=\{i : y_{i}=h_y\}$. In a few words, \eqref{eq::cond} is
 equivalent to have at least one absence and one presence of drug consumption
 in both sets $\{x_{ij}\mid y_i = 0\}$ and $\{x_{ij}\mid y_i = 1\}$. To ensure
 that the MLE is well defined, this condition suggests us to
 do not take into account drugs that do not satisfy it. 

\section{Model selection by MCMC algorithm} \label{sec::MH}
\subsection{Bayesian model selection}
We define the set of the competing models $\Gamma$ as the set of models
$\bgamma \in\mathbb{B}^p$ where \eqref{eq::cond} is satisfied. So,
\begin{equation}
  \Gamma=\{ \bgamma\in\mathbb{B}^p \text{ such as } \eqref{eq::cond}
  \text{ is satisfied for }   \bgamma \}. \label{eq::gamma}
\end{equation}

In a Bayesian framework, the aim is to obtain the model having the
highest posterior distribution $p\big(\bgamma \mid \by, \bX \big)$. We
assume that uniformity holds for the prior distribution $p(\bgamma
\mid \bX)$ of models $\bgamma \in \Gamma$. So, we have
\begin{equation}
p\big(\bgamma \mid \by, \bX \big) \propto p(\by \mid  \bX, \bgamma),
\end{equation}
where $p(\by \mid  \bX, \bgamma)$ is the \emph{integrated likelihood} defined by
\begin{equation}
p(\by \mid  \bX, \bgamma) = \int_{\bOG} p(\by \mid  \bX, \bgamma, \bbeta) p(\bbeta \mid  \bX, \bgamma)d\bbeta,
\end{equation}
where $p(\by \mid \bX, \bgamma, \bbeta)=\exp\big(\ell_n(\by \mid \bX,
\bgamma, \bbeta)\big)$ is the likelihood related to model $\bgamma$ and where
$p(\bbeta \mid \bX, \bgamma)$ is the prior distribution of $\bbeta$
whose the support is included in $\bOG$. Since logarithm is monotone,
\begin{equation}
\argmax_{\bgamma \in\Gamma} p\big(\bgamma \mid \by, \bX \big) = \argmax_{\bgamma \in \Gamma} \ln p(\by \mid  \bX, \bgamma).
\end{equation}

When the integrated likelihood has not a closed form, the Bayesian
Information Criterion (BIC) is generally used. It is based on a second
degree Laplace approximation of the logarithm of the integrated
likelihood \citep{Sch78}, and it is defined as
\begin{equation}
\label{eq:bic}
 \BIC(\bgamma) = \ell_n\big(\by\mid \bX, \bgamma,  \widehat{\bbeta}_{\bgamma}\big)
 - \dfrac{\nu_{\bgamma}}{2} \ln n,  
\end{equation}
where $\nu_{\bgamma}=1+\sum_{j=1}^p \gamma_j$ is the degree of freedom for model $\bgamma$. Therefore, we want to
achieve $\bgammastar$ which is the model maximizing the BIC criterion, so
\begin{equation}
\bgammastar = \argmax_{\bgamma \in \Gamma} \BIC(\bgamma).
\end{equation}
This criterion selects the model providing the best trade-off between
its accuracy related to the data and its complexity.

Obviously, the number of competing models is too huge for applying an exhaustive approach (\emph{i.e.} to compute the BIC criterion for each model). Therefore, the Metropolis-Hastings
algorithm described in the following section is used to estimate $\bgammastar$.

\subsection{Metropolis-Hastings algorithm for achieving $\bgammastar$}
Model $\bgammastar$ can be achieved through a Metropolis-Hastings algorithm
\citep{Rob04}, described in Algorithm~\ref{algo:MH}, which performs a random
walk over $\Gamma$. The unique invariant distribution of
Algorithm~\ref{algo:MH} is proportional to $\exp{\big(\BIC(\bgamma)\big)}$.
Therefore, $\bgammastar$ is the mode of its stationary distribution.

At each iteration, the algorithm proposes to move into a neighbourhood of the
current model. A neighbouring model is defined as copy of the current model
where just a few elements are altered. Thus, at iteration $[r]$, the candidate
$\tilde{\bgamma}$ is equal to the current model $\bgamma^{[r]}$ except for
$\alpha\geq 1$ elements at the maximum. More specifically, $\tilde{\gamma}$ is
uniformly sampled in $V_{\alpha}(\bgamma^{[r]})$ where
\begin{equation}
V_{\alpha}(\bgamma^{[r]}) = \left\{
\bgamma: 
\sum_{j=1}^p |\gamma_j - \gamma_j^{[r]}|\leq\alpha
\right\}.
\end{equation}
In the application, we set $\alpha=5$ to obtain good mixing properties. The candidate $\tilde{\bgamma}$ is
accepted with a probability equal to
\begin{equation}
\rho^{[r]}=\frac{\exp\big(\BIC(\tilde{\bgamma})\big)}{\exp\big(\BIC(\bgamma^{[r]})\big)}.
\end{equation}
Note that we define that $\BIC(\bgamma)=-\infty$ for all
$\bgamma\in\mathbb{B}^{p}\setminus\Gamma$. This algorithm performs $R$
iterations and returns the model maximizing the BIC criterion. In practice,
there may be almost absorbing states, so different
initialisations of this algorithm ensure to visit $\bgammastar$.

\begin{algorithm}[h!]
\caption{Metropolis-Hasting performing the model selection} \label{algo:MH}
%\begin{algorithmic}
  \textbf{Initialisation} $\bgamma^{[0]}$ is uniformly sampled in $\Gamma$. \\
   \textbf{For} $r=1,\ldots,R$.\\
	\hspace*{0.15cm}  \textbf{Candidate step:} $\tilde{\bgamma}$ is uniformly sampled in $V_{\alpha}(\bgamma^{[r]})$.
  
  \hspace*{0.15cm} \textbf{Acceptance/reject step:} defined $\bgamma^{[r]}$ with
   	\begin{equation*}
 	\bgamma^{[r]} =\left\{ \begin{array}{rl}
 	\tilde{\bgamma} & \text{with probability } \rho^{[r]} \\
 	\bgamma^{[r-1]} & \text{otherwise}
 	\end{array}\right. .
 	\end{equation*}
\textbf{End For}\\
\textbf{Return} $\displaystyle{\argmax_{r=1,\ldots,R} \BIC(\bgamma^{[r]})}$.
%\end{algorithmic}
\end{algorithm}

\section{Results on real data set} \label{sec::application} 
In this section, after presenting the French pharmacovigilance database, the proposed method is
compared to the others by using the OMOP set. Finally, specific comments are given for the proposed method.

\subsection{Data}
To evaluate and compare the performances of the competing methods, we use the
OMOP~\citep{Rya13} reference set of test cases that contains both
positive and negative controls.  Four adverse events (\emph{i.e.} $d=4$) were studied in
this reference set : acute myocardial infarction (AMI), acute kidney
injury (AKI), acute liver injury (ALI), and upper gastro-intestinal
bleeding (GIB). There are three-hundred and ninety-nine test cases
where $165$ positive controls and $234$ negative controls were
identified across the four adverse events of interest. More details are given by Table~\ref{tab::omop1}.   \cite{Rya13}
indicate that the majority of positive controls for AKI and GIB were
supported by randomized clinical trial evidence, while the majority of
positive controls for ALI and AMI were only based on published case
reports.

\begin{table}[htp]
\begin{center}
\caption{Numbers of positive and negative controls for the four adverse event in the OMOP reference set.}\label{tab::omop1} 
\begin{tabular}{ccccc}
  control & AMI & GIB & ALI & AKI \\ 
\hline  positive & 36 & 24 & 81 & 24 \\
 negative & 66 & 67 & 37 & 64 \\
  \hline
\end{tabular}
\end{center}
\end{table}

 Methods are compared on the data extracted from the French
pharmacovigilance database where notifications have been collected
from 2000 to 2010. The studied database contains $n=219,340$
individuals notifications and the consumption informations concerning
$p=145$ drugs mentioned on the OMOP reference set. Therefore, $145\times4=580$ drug-event pairs are studied, among them 145 are positive controls ($25\%$), 153
are negative controls ($26\%$) and 282 have an unknown status
($49\%$). More details are given in Table~\ref{tab::omop2}.  The four studied adverse events occur 495 (AMI), 4746 (GIB), 10910 (ALI) and 5234 (AKI) times in the French pharmacovigilance database.

\begin{table}[htp]
\begin{center}
\caption{Numbers of positive, negative and unknown signals for the four adverse event in the OMOP reference set and for the 145 drugs presented in both databases (OMOP and French pharmacovigilance).}\label{tab::omop2} 
\begin{tabular}{ccccc}
  control & AMI & GIB & ALI & AKI \\ 
\hline  positive & 29 & 20 & 75 & 21 \\
 negative & 43 & 46 & 22 & 42 \\
 unknown & 73 & 79 & 48 & 82 \\
  \hline
\end{tabular}
\end{center}
\end{table}

\subsection{Competing methods}
\paragraph{Disproportionality-based methods}
We chose to compare our method with all the disproportionality methods implemented in the R package \emph{PhViD} \citep{Ahm13} Thus, four disproportionality-based methods: the Proportional Reporting Ratio (PRR),
the Reporting Odds Ratio (ROR), the Reporting Fisher Exact Test (RFET) \citep{Ahm10} and the
FDR-based Gamma Poisson Shrinkage (GPS) \citep{Ahm09} are considered. The specific statistics are used with a threshold of 0.05 and are presented in Table~\ref{tab::stat}. All methods are compared
on the $580$ drug-event pairs mentioned on the OMOP reference set.
\begin{table}[htp]
\begin{center}
\caption{Specific statistics of the disproportionality methods: statistics (Stat), minimal number of individuals having a drug-event pair to claim this pair as a signal (Min.) and reference (Ref.).}\label{tab::stat} 
\begin{tabular}{cccc}
  Method & Stat. & Min. & Ref. \\ 
\hline  PRR & p-value of rank & 3 & \citet{Eva01} \\ 
  ROR & p-value of rank & 3 & \citet{Van02} \\ 
  RFET & mid-pvalue & 1 & \citet{Ahm10}\\ 
  GPS & prob of H0 & 1 &  \citet{Ahm09}\\ 
  \hline
\end{tabular}
\end{center}
\end{table}

\paragraph{Lasso-based logistic regressions}
The results of the Lasso method applied on logistic regressions are obtained
with the R package \emph{glmnet} \citep{Fri10}. The penalty value is selected
by cross-validation with ten folds to obtain the most parsimonious model among
the models having best misclassification error. This method permits to find
few signals since the selected penalty implies that only the intercept is not
zero for only one adverse event (AMI). This example shows the difficulty for calibrating the Lasso-penalty. Indeed, the misclassification error is
roughly constant according to the penalty value. This is due to the weak rate of
notifications for one adverse event.

\paragraph{Model-based logistic regressions}
For each of the four adverse events, 100 random initialisations of
Algorithm~\ref{algo:MH} with $\alpha=5$ and $R=5.10^3$ iterations have been
done. The model maximizing the BIC criterion is returned. Table~\ref{tab::MH}
presents the number of competing models for each adverse event, which corresponds to the dimension of $\Gamma$ defined in \eqref{eq::gamma}.

\begin{table}[ht!]
\begin{center}
\caption{Number of drugs respecting \eqref{eq::cond} and number of competing models for each adverse event ($| \Gamma |$).}\label{tab::MH}
\begin{tabular}{ccccc}
Adverse Event & AMI & GIB & ALI & AKI \\ 
\hline Number of  drugs respecting \eqref{eq::cond} & 66 & 97 & 123 & 107 \\
 $| \Gamma |$ & $2^{66}$ & $2^{97}$ & $2^{123}$ & $2^{107}$ \\
 \hline
\end{tabular} 
\end{center}
\end{table}

\subsection{Method comparison}
Table~\ref{tab::main_results} presents the rates of positive controls,
of negative controls and of unknown signals detected by all the
competing methods.

\begin{table}[ht!]
\begin{center}
\caption{Main results obtained by the competing methods ordered by their rate of positive controls: number of signals (NS), rate of positive controls (RPC), rate of negative controls (RNC) and rate of unknown signals (RUS).}\label{tab::main_results}
\begin{tabular}{ccccc}
  Method & NS & RPC & RNC & RUS \\ 
  \hline Logistic BIC (Algorithm~\ref{algo:MH})& 70 & 0.54 & 0.01 & 0.45 \\
  RFET & 114 & 0.51 & 0.06 & 0.43 \\  
  PRR  & 73 & 0.51 & 0.10 & 0.40 \\ 
  ROR & 120 & 0.50 & 0.07 & 0.43 \\ 
  GPS  & 129 & 0.48 & 0.07 & 0.45 \\ 
  Lasso-CV & 13 & 0.46 & 0.08 & 0.46 \\ 
  \hline 
\end{tabular} 
\end{center}
\end{table}

The proposed method obtains the best rates of positive controls and
negative controls. It detects $70$ signals while the Lasso-based method
finds only 13 couples. The poor results of the Lasso are explained by
the penalty values assessed by the misclassification error rate.  Indeed, the resulting penalty values constrain all the
coefficients to be zero for three adverse events. 
All the disproportionality methods obtain similar results.
 Despite that many signals are detected by these methods (between 73 and 129), their rates of positive and negative controls are worse than those resulting from the proposed method.
 
 Since the proposed method obtains the best rates of positive and negative controls, we conclude that it is more precise for the signal detection. However, it finds less signals than the disproportionality methods. So, it permits the practitioner to focus on more probably related drug-event pairs. Moreover, some associations detected only by the disproportionality method could be due to the co-prescription phenomenon.
 
\subsection{Specific comments about the proposed method}

Table~\ref{tab::temps} indicates the computing time obtained by an Intel(R)
Xeon(R) CPU 3.00 GHz and the number of times where the Algorithm~\ref{algo:MH} finds the
best model.

\begin{table}[ht!]
\begin{center}
\caption{General results of Algorithm~\ref{algo:MH}: number of times where
  $\bgammastar$ has been found (model), number of signals (nb signals), number
  of positive controls, number of negative controls, computing time in minutes
  required for one Markov chain realization (time) and number of unique profiles for the best
  model ($m_{\bgammastar}$).}\label{tab::temps}
\begin{tabular}{ccccc}
Adverse Event & AMI & GIB & ALI & AKI \\ 
\hline 
 model & 100  &   67  &   50  &   56  \\
 nb signals &  9  &   10  &   26  &   25    \\
 positive controls &  1  &   5  &   20  &   12    \\
 negative controls &  1  &   0  &   0  &   0    \\
 time &  1  &   3  &   3  &   5    \\
 $m_{\bgammastar}$ &  45  &   629  &   554  &   1024  \\
\hline
\end{tabular} 
\end{center}
\end{table}

The computing time has been strongly reduced by using the expression of the
log-likelihood given in Appendix A. For example, by considering the best model
resulting of the adverse event AMI where 9 variables have a non-zero coefficient,
the database can be reduced to $m_{\bgammastar}=45$
unique weighted individuals (see Appendix~A). Moreover, since many
different initialisations allow to find $\bgammastar$, the number of
initialisations (set at 100 during the experiment) could be reduced.
 Finally, the list of the detected signals are presented in Appendix~B.

\subsection{Specific comments about the Lasso}
We have seen that the Lasso obtains poor results when the penalty is
determined according to the misclassification error. \cite{Cas10} suggest to
set the same penalty value for all the adverse events. Moreover, they use a
disproportionality measure to evaluate the number of signals and thus to
deduce the penalty value.

\begin{figure}[ht!]
\begin{center}\caption{Rates of positive and negative controls obtained by the Lasso with different penalities black curve) and obtained by the model maximizing the BIC criterion (red dots).} 
\label{fig::Lasso}
\includegraphics[scale=0.5]{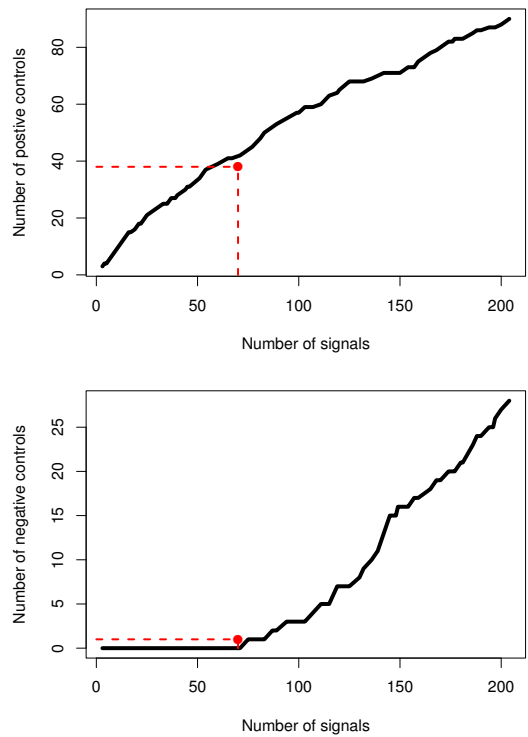}
\end{center}
\end{figure}

In order to investigate the Lasso approach features, we build a sequence of
penalties to obtain different numbers of signals with the Lasso. The numbers
of positive and negative controls resulting for each penalty values are
indicated by the black lines of Figure~\ref{fig::Lasso}.

The results related to the model maximizing the BIC criterion are indicated by
 red dots. On Figure~\ref{fig::Lasso}, it is very hard to find a penalty
value from where results obtained a better trade off between the positive and
the negative controls. If, for the same number of signals ($70$) as obtained by
Algorithm~\ref{algo:MH}, the Lasso approach presents slightly better
performances, the corresponding penalty value does not result from an
optimizing procedure. These figures can not be plotted in reality, since the
nature of the signals are unknown.
Thus, it seems more efficient to select the model maximizing the BIC criterion than to use a Lasso regression. Indeed, the penalty calibration is very difficult and the results related to the ''best'' penalty value are similar to those related to the model maximizing the BIC criterion. Moreover, this penalty value is not accessible in practice.

\section{Discussion} \label{sec::discussion}

In this paper, we have proposed a method for analysing individual spontaneous
reporting databases, which also avoids the drawbacks of the
disproportionality-based measures (co-prescription and masking effects). The signal detection is
led throughout parsimonious logistic regressions whose sparsity degree is
assessed as a model selection challenge. Therefore, we avoid the use of
Lasso-type method that requires the challenging calibration of penalty.
%Indeed, in pharmacovigilance settings, the cross-validated missclassification
%error fails to set an efficient penalty value. 
The combinatorial problem of
model selection is bypassed by Metropolis-Hastings binary space sampling.

Despite to the difficulties for evaluating pharmacovigilance methods, the OMOP
reference set of \cite{Rya13} gives us the opportunity to compare the proposed method
to the reference approaches on real data. On these data, it
appears to be relevant for the signal detection issue. However, many
signals are not detected by our method. So, we conclude that this method
should be used in parallel to existing measures in pharmacovigilance.

\begin{figure}[ht!]
\begin{center}
\caption{Evolution of the number of drugs verifying \eqref{eq::cond} according to the headcount of the adverse event.} \label{fig::dim}
\includegraphics[scale=0.5]{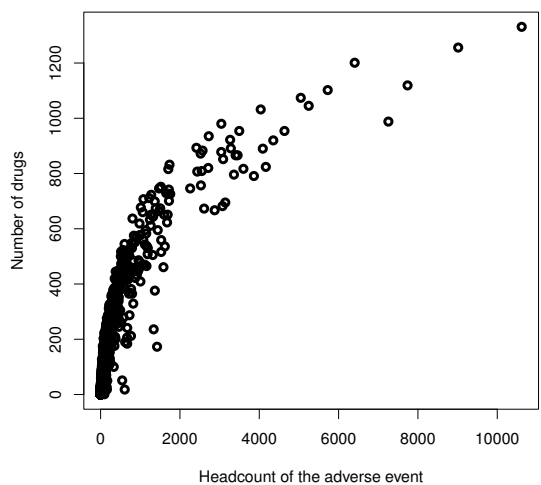}
\end{center}
\end{figure}

The proposed approach can manage the whole French pharmacovigilance
database which consists of $n=219,340$ individual notifications,
$p=2,114$ drugs and $d=4,257$ adverse events. We have shown that the
dimension of the model space is defined by the number of drugs
verifying \eqref{eq::cond}.  Figure~\ref{fig::dim} presents the
evolution of this number according to the headcount of the adverse
events.

In the whole database, $75\%$ of the adverse events can be associated to less
than 42 drugs. For the adverse events which have less than 12 drugs verifying
\eqref{eq::cond}, we advise to use an exhaustive approach consisting of
computing the BIC criterion for each competing models in $\Gamma$. The model
selection on the whole French pharmacovigilance database is achieved at the
cost of several days of computing time. The proposed approach can thus be used to
investigate targeted adverse events. Finally, a preliminary drug selection
could provide a reducing of computing time.

\begin{acknowledgement}
The authors are grateful to Isma\"{i}l Ahmed for his
useful advices. This work was supported by the French agency for
drug safety.
\end{acknowledgement}

%\begin{thebibliography}{10}
%\bibitem[Author(2000)]{aa} xxx.
%\end{thebibliography}
\bibliographystyle{natbib} 
\bibliography{biblio}
\appendix

\section{Weighted form of the adverse event log-likelihood} \label{sec::weighted}

Obviously, the coordinates of $\bx_i$ impacting the log-likelihood
value are those belonging to $\mathcal{D}_{\bgamma}$. For each
observation $\bx_i$, we denote by
$\bx_i^{\bgamma}\in\mathbb{B}^{|\bgamma|}$, where
$|\bgamma|=\sum_{j=1}^p \gamma_j$, the vector containing the elements
of $\bx_i$ impacting the log-likelihood (\emph{i.e.} the vector
composed with the elements of $\bx_i$ such as index belongs to
$\mathcal{D}_{\bgamma}$). Thus, for each $j=1,\ldots,|\bgamma|$:
\begin{equation}
x_{ij}^{\bgamma}=x_{ij_0} \text{ with } j_0=\min\left\{j': \sum_{j''=1}^{j'}\gamma_{j''}=j\right\}.
\end{equation}  
  
Moreover, many individual profiles $(\bx_i^{\bgamma},y_i)$ occur many
times in the database. We denote by $m_{\bgamma}$ the number of
different profiles impacting the log-likelihood of model
$\bgamma$. The profile $i$ is denoted by
$(\tilde{\bx}_i^{\bgamma},\tilde{y}_i^{\bgamma})$ and its weight is
denoted by $w_i^{\bgamma}$. Thus, \eqref{eq:log-lik} is given by
  \begin{align}
  \label{eq:w-log-lik}
    \ell_n\big(\by\mid \bX, \bgamma, \bbeta \big) = & \sum^{m_{\bgamma}}_{i = 1}
    w_i^{\bgamma} \tilde{y}_i^{\bgamma} \big(\beta_0 + \sum_{j =1}^{|\bgamma|}\beta_j
    \tilde{x}_{ij}^{\bgamma}\big)\notag\\
    & - w_i \ln \Big[1 + \exp\big(\beta_0 + \sum_{j=1}^{| \bgamma|}\beta_j \tilde{x}_{ij}^{\bgamma}\big)\Big],
  \end{align}
  where $\beta_j^{\bgamma}$ is the $j$-th element which is not zero in $\bbeta$, so for each $j=1,\ldots,|\bgamma|$:
\begin{equation}
\beta_{j}^{\bgamma}=\beta_{j_0} \text{ with } j_0=\min\left\{j': \sum_{j''=1}^{j'}\gamma_{j''}=j\right\}.
\end{equation}  
  
In practice, it is often more numerically efficient to compute the
adverse event log-likelihood by using ~\eqref{eq:w-log-lik} than by
using \eqref{eq:log-lik}.

\section{Signals detected by the proposed methods}\label{sec::signals}
Table~\ref{tab::signals} presents the couples between a drug and an adverse events  detected by the proposed method.

\begin{table}[ht!]
\begin{center}
\caption{List of the signals detected by the proposed method.}\label{tab::signals}
\begin{scriptsize}
\begin{tabular}{ccccc}
 Adverse event   &  ATC  & Headcount &  $\beta_j$   &  Omop control  \\ 
\hline AMI&L03AB07&7&2.92&unknown\\
ALI&J05AE09&36&2.85&positive\\
AMI&N02CC03&6&2.6&positive\\
ALI&L01BB03&8&2.51&positive\\
AKI&M01AE09&46&2.27&unknown\\
ALI&C02KX01&65&2.25&positive\\
AMI&M01AH01&21&2.05&unknown\\
AMI&L01BC05&10&1.85&unknown\\
GIB&M01AC01&138&1.84&positive\\
AMI&J05AF05&92&1.84&unknown\\
GIB&B01AC04&523&1.77&positive\\
AKI&J05AF07&144&1.76&unknown\\
GIB&B01AC07&31&1.67&unknown\\
AKI&C09AA05&353&1.66&unknown\\
AKI&C09AA03&165&1.66&positive\\
AKI&C09CA08&35&1.65&positive\\
ALI&L02BB01&10&1.61&positive\\
ALI&J05AG01&297&1.55&positive\\
ALI&J02AC03&117&1.54&positive\\
AKI&C09AA02&146&1.51&positive\\
GIB&M01AE03&276&1.5&positive\\
AKI&C09AA10&38&1.49&unknown\\
AKI&N05AD08&10&1.48&unknown\\
AKI&L04AD01&91&1.38&positive\\
GIB&M01AE02&52&1.34&positive\\
ALI&J01XE01&52&1.33&positive\\
ALI&J04AB02&538&1.31&positive\\
AKI&C09CA07&34&1.31&positive\\
AKI&M01AE03&250&1.31&positive\\
AMI&L04AB02&13&1.29&unknown\\
AKI&L01BA01&129&1.28&unknown\\
ALI&A03AX13&26&1.25&unknown\\
ALI&A07EC01&71&1.24&unknown\\
AKI&L01BC05&57&1.18&unknown\\
AKI&C09CA06&139&1.16&positive\\
GIB&M01AH01&98&1.15&unknown\\
ALI&J02AC02&22&1.15&positive\\
AMI&B01AC04&24&1.14&unknown\\
ALI&J04AC01&359&1.08&positive\\
AKI&C09AA06&25&1.06&unknown\\
AKI&C09AA01&61&1.02&positive\\
ALI&N03AF01&248&0.99&positive\\
AKI&M01AE02&43&0.99&positive\\
ALI&D01AE15&77&0.98&positive\\
AMI&J05AF02&30&0.98&negative\\
ALI&L03AB07&27&0.98&positive\\
ALI&J02AC01&188&0.97&positive\\
ALI&G03CA03&76&0.96&unknown\\
AKI&J04AB02&104&0.96&unknown\\
GIB&A12BA01&155&0.94&positive\\
AKI&J01MA02&147&0.87&unknown\\
AMI&J05AF06&44&0.83&unknown\\
ALI&L01BA01&186&0.81&positive\\
AKI&M04AA01&220&0.77&positive\\
AKI&C03AA03&430&0.73&positive\\
AKI&M01AH01&72&0.68&unknown\\
GIB&C08DB01&81&0.63&unknown\\
AKI&A12BA01&154&0.62&unknown\\
ALI&A10BF01&36&0.61&unknown\\
AKI&M01AC01&47&0.59&positive\\
ALI&N03AG01&298&0.56&positive\\
ALI&J01MA06&60&0.54&positive\\
ALI&N05BA05&147&0.53&unknown\\
ALI&J05AF07&177&0.49&positive\\
ALI&M04AA01&216&0.45&positive\\
AKI&J01MA01&109&0.43&unknown\\
GIB&C08CA01&126&0.37&unknown\\
ALI&N06AB04&117&0.36&unknown\\
GIB&C09AA05&148&0.35&unknown\\
ALI&M01AE03&200&0.31&unknown\\
ALI&J01MA02&202&0.31&positive\\
\end{tabular}
\end{scriptsize}
\end{center}
\end{table}
\end{document}